\documentclass[prl,aps,amssymb,superscriptaddress,twocolumn,10pt]{revtex4}

\usepackage{amssymb}
\usepackage{graphicx}

\def\og{\leavevmode\raise.3ex\hbox{$\scriptscriptstyle\langle\!\langle$~}}
\def\fg{\leavevmode\raise.3ex\hbox{~$\!\scriptscriptstyle\,\rangle\!\rangle$}}

\newcommand{\unit}[1]{\mathrm{~#1}}
\newcommand{\guil}[1]{``#1''}
\newcommand{\moy}[1]{\left < #1 \right >}
\newcommand{\inp}[1]{\left(#1\right)}
\newcommand{\insb}[1]{\left[#1\right]}
\newcommand{\incb}[1]{\left\{#1\right\}}
\newcommand{\abs}[1]{\left|#1\right|}

\newcommand{\ee}[1]{\times 10^{#1}}
\newcommand{\vdc}{V_{\mathrm{dc}}}
\newcommand{\vac}{V_{\mathrm{ac}}}
\newcommand{\frPerm}{Reprinted figure with permission from J.-C. {Forgues}, C.~{Lupien}, B.~{Reulet}, Physical Review Letters, 114, 130403, 2015. \copyright 2015 by the American Physical Society.}

\begin{document}

\title{Non-classical radiation emission by a coherent conductor}

\author{Jean-Charles Forgues}\email{jean-charles.forgues@usherbrooke.ca}
\author{Christian Lupien}\email{christian.lupien@usherbrooke.ca}
\author{Bertrand Reulet}\email{bertrand.reulet@usherbrooke.ca}
\affiliation{D\'{e}partement de Physique, Universit\'{e} de Sherbrooke, Sherbrooke, Qu\'{e}bec, Canada, J1K 2R1}

\date{\today}
\begin{abstract}
We report experimental evidence that the microwave electromagnetic field generated by a normal conductor, here a tunnel junction placed at ultra-low temperature, can be non-classical. By measuring the quadratures of the electromagnetic field at one or two frequencies in the GHz range, we demonstrate the existence of squeezing as well as entanglement in such radiation. In one experiment, we observe that the variance of one quadrature of the photo-assisted noise generated by the junction goes below its vacuum level. In the second experiment, we demonstrate the existence of correlations between the quadratures taken at two frequencies, which can be stronger than allowed by classical mechanics, proving that the radiation at those two frequencies are entangled.
{\it To cite this article: J.-C. Forgues, G. Gasse, C. Lupien, B. Reulet, C. R. Physique 6 (2016).} (Inspired from previous works/Inspir\'e de travaux ant\'erieurs \cite{Gasse2013,Forgues2015})

\vskip 0.5\baselineskip
\noindent{\bf Rayonnement non classique \'emis par un conducteur coh\'erent}
Nous rapportons des preuves exp\'erimentales que le champ \'electromagn\'etique micro-ondes g\'en\'er\'e par un conducteur normal, une jonction tunnel plac\'ee \`a ultra-basse temp\'erature, peut avoir un comportement non-classique. Nous d\'emontrons l'existence de compression d'\'etat ainsi que d'enchev\^etrement dans cette radiation en mesurant les quadratures du champ \'electromagn\'etique \`a une ou deux fr\'equences de l'ordre du GHz. Dans une exp\'erience, nous observons que la variance d'une quadrature du bruit photo-assist\'e g\'en\'er\'e par la jonction descend sous son niveau du vide. Dans une deuxi\`eme exp\'erience, nous d\'emontrons l'existence de corr\'elations entre les quadratures observ\'ees \`a deux fr\'equences, corr\'elations qui peuvent \^etre sup\'erieures \`a ce qui est permis par la m\'ecanique classique, prouvant que la radiation \`a ces deux fr\'equences est enchev\^etr\'ee.
{\it Pour citer cet article~: J.-C. Forgues, G. Gasse, C. Lupien, B. Reulet, C. R. Physique 6 (2016).}

\noindent{\small{\it Keywords~:} Quantum Microwaves; Entanglement; Squeezing; Quantum Noise; Shot Noise; Tunnel Junction}
\vskip 0.5\baselineskip
\noindent{\small{\it Mots-cl\'es~:} Micro-ondes quantiques~; Enchev\^etrement~; Compression d'\'etat~; Bruit Quantique~; Bruit de grenaille~; Jonction tunnel}
\end{abstract}

\maketitle

\section{Introduction}
\label{secIntro}

A great effort is currently deployed to find sources of quantum light. A light with properties beyond that of classical physics is indeed essential to the development of quantum information technology \cite{Braunstein2005,Weedbrook2012} and has direct applications in metrology \cite{Caves1981}. Quantum light can be non-classical in several ways. Squeezed light offers the possibility of observing fluctuations lower than that of vacuum along one quadrature: the rms flutuations $\Delta X^2$ of the (in phase) amplitude of $X\cos\inp{2\pi f_1t}$ can be smaller than that of vacuum at the expense of an increase of the rms fluctuations $\Delta P^2$ of the (quadrature) amplitude of $P\sin\inp{2\pi f_1t}$; this is necessary in order to preserve Heisenberg's uncertainty principle (for a review on squeezing, see \cite{Loudon2000,Gardiner2004,Wiseman2009}). Two-mode squeezed light refers to the existence of correlations between the quadratures of the electromagnetic field taken at two different frequencies $f_1$ and $f_2$ that go beyond what is allowed by classical mechanics\cite{Forgues2015}. A strong enough two-mode squeezing can lead to entanglement between the two frequencies\cite{Duan2000}.

Many systems have been devised to produce squeezed light, based for example on non-linear crystals, atomic transitions and non-linear cavities in optics \cite{Slusher1985}, but also with parametric amplifiers and qubits in the microwave domain \cite{Yurke1988,Movshovich1990,Nation2012,Flurin2012}. The key ingredient in all these systems is the existence of a nonlinearity, which allows the mixing of vacuum fluctuations with the classical, large field of a coherent pump. Here we use the discreteness of the electron charge $e$ as a source of non-linearity: A tunnel junction (two metallic contacts separated by a thin insulating layer) has \emph{linear} $I(V)$ characteristics at low voltage and thus cannot be used as a non-linear element to mix signals. There is no photo-assisted dc transport, i.e. no rectification. However, electrical current $I(t)$ flowing in a conductor always fluctuates in time, a phenomenon usually referred to as \guil{electrical noise}. Interestingly, this noise can be \emph{non-linear} as a function of voltage, even if the $I(V)$ characteristics itself is linear.

While the dc current corresponds to the average $\moy{I(t)}$, current fluctuations are characterized by their statistical properties such as their second order correlator $\moy{I\inp tI\inp{t'}}$ or, in frequency space, the noise spectral density $S\inp f=\moy{\abs{I\inp f}^2}$ where $I \inp f$ is the Fourier component of the current at frequency $f$. Here the brackets $\moy{\cdots}$ represent the statistical average. The tunnel junction, as well as most coherent conductors, exhibits shot noise: the variance $\Delta I^2$ of the current fluctuations  generated by the junction depends on the bias voltage. For example, at low frequency and high current, the noise spectral density is given by $S\inp{f_1=0}=e\abs{I}$ (for a review on shot noise in mesoscopic conductors, see \cite{Blanter2000,Nazarov2003}), a strongly non-linear function. When under ac excitation, the junction exhibits photo-assisted noise \cite{Lesovik1994,Schoelkopf1998,Kozhevnikov2000} as well as a dynamical modulation of its noise \cite{Gabelli2008,Gabelli}. We use this modulation of the \emph{intrinsic} noise of the junction by an external ac excitation to generate squeezing.

An alternate approach is to consider that the time-dependent current fluctuations in the sample generate a random electromagnetic field that propagates along the electrical wires. Both these descriptions are equivalent. For example, the equilibrium current fluctuations (Johnson-Nyquist noise \cite{Johnson1928,Nyquist1928}) correspond to the blackbody radiation in one dimension\cite{Oliver1965}. More precisely, the power radiated by a sample at frequency $f$ in a cable is proportional to the spectral density $S\inp f$ of current fluctuations which, at high temperature and at equilibrium (i.e. with no bias), is given by $S\inp{hf\ll k_BT}=2k_BT/R$ where $T$ is the temperature and $R$ the electrical resistance of the sample\cite{Callen1951}. 

In short samples at very low temperatures, electrons obey quantum mechanics. Thus, electron transport can no longer be modeled by a time-dependent, classical number $I\inp t$, but needs to be described by an operator $\hat I\inp t$. Current fluctuations are characterized by correlators such as $\moy{\hat I\inp t\hat I\inp{t'}}$. Quantum predictions differ from classical ones only when the energy $hf$ associated with the electromagnetic field is comparable with energies associated with the temperature $k_BT$ and the voltage $eV$. Hence for $hf\gg k_BT,eV$, the thermal energy $k_BT$ in the expression of $S\inp f$ has to be replaced by that of vacuum fluctuations, $hf/2$. Some general link between the statistics of current fluctuations and that of the detected electromagnetic field is required beyond the correspondence between spectral density of current fluctuations and radiated power \cite{Beenaker2001,Beenaker2004,Lebedev2010,Grimsmo2016,Mendes2015}. In particular, since the statistics of current fluctuations can be tailored by engineering the shape of the time-dependent bias voltage \cite{Gabelli2013}, it is possible to induce non-classical correlations in the electromagnetic field generated by a quantum conductor. For example, an ac bias at frequency $f_0$ generates correlations between current fluctuations at frequencies $f_1$ and $f_2$, i.e. $\moy {\hat I(f_1)\hat I(f_2)}\neq 0$, if $f_1\pm f_2=nf_0$ with $n$, an integer \cite{Gabelli2008,Gabelli,Gabelli2007}. This is responsible for the existence of correlated power fluctuations \cite{C4classique} and for the emission of photon pairs \cite{Forgues2014} recently observed. For $f_1=f_2$, $\moy{\hat I^2(f_1)}\neq 0$ leads to vacuum squeezing .

Entanglement of photons of different frequencies has already been observed in superconducting devices engineered for that purpose in Refs. \cite{Eichler2011a,Flurin2012,Nguyen2012}, where frequencies $f_1$ and $f_2$ are fixed by resonators and the entanglement comes from a non-linear element, a Josephson junction. What we show here is that a quantum conductor excited at frequency $f_0$ can emit entangled radiation at \emph{any} pair of frequencies $f_1$, $f_2$ such that $f_0=f_1+f_2$. This property is demonstrated using a tunnel junction but our results clearly stand for any device that exhibits quantum shot noise.
The key ingredient for the appearance of entanglement is the following: noise at any frequency $f_1$ modulated by an ac voltage at frequency $f_0$ gives rise to sidebands with a well-defined phase. These sidebands, located at frequencies $\pm f_1\pm nf_0$ with $n$, an integer, are correlated with the current fluctuations at frequency $f_0$. The particular case $f_2=-f_1+f_0$ we study here corresponds to the maximum correlation.

In this article, we report two measurements that exhibit quantum properties of the electromagnetic field generated by a tunnel junction under ac excitation. First, we use an excitation at frequency $f_0=f_1$ or $f_0=2f_1$ to induce a non-zero correlator $\moy{\hat I(f_1)^2}$, responsible for the existence of radiation squeezing at frequency $f_1$\cite{Bednorz2013,Gasse2013}. Second, we induce correlations between electromagnetic field quadratures at two frequencies $f_1$ and $f_2$ by irradiating the sample at frequency $f_0=f_1+f_2$.  By analyzing these correlations, we show that the electromagnetic field produced by electronic shot noise can be described in a way similar to an Einstein-Podolski-Rosen (EPR) photon pair: when measuring fluctuations at only one frequency, i.e. one mode of the electromagnetic field, no quadrature is preferred. But when measuring two-modes, we observe strong correlations between identical quadratures. These correlations are stronger than what is allowed by classical mechanics as proven by their violation of Bell-like inequalities.

\section{Experimental setup \cite{fn1}}
\begin{figure}
    \centerline{ \includegraphics[width=\columnwidth] {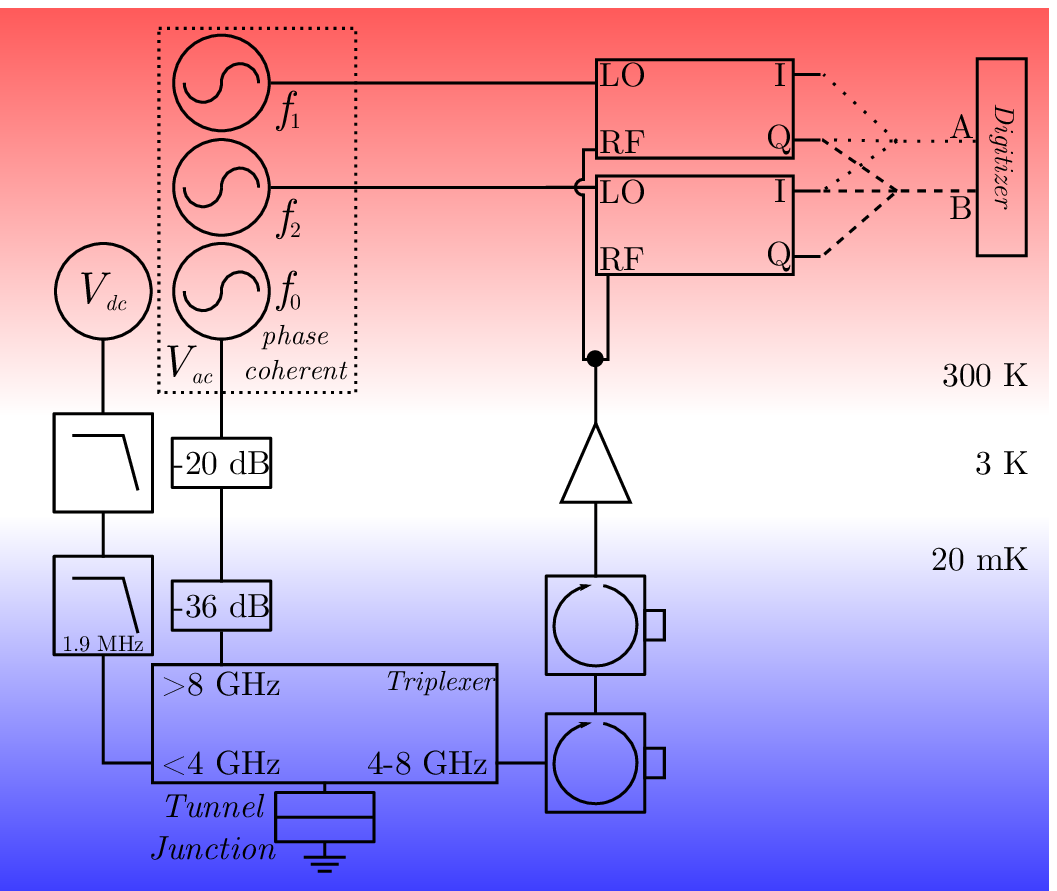}}
    \caption{\footnotesize (colour online) Experimental set-up. See details in text./Montage exp\'erimental. Voir le texte pour les d\'etails. \frPerm}
\label{figMontage}
\end{figure}

In Fig. \ref{figMontage}, we use a $R=70\unit{\Omega}$ Al/Al$_2$O$_3$/Al tunnel junction in the presence of a magnetic field to insure that the aluminium remains a normal metal at all temperatures. It is cooled down to $\sim18\unit{mK}$ in a dilution refrigerator. A triplexer connected to the junction separates the frequency spectrum in three bands corresponding to the dc bias ($<4\unit{GHz}$), the ac bias at frequency $f_0$ ($>8\unit{GHz}$) and the detection band ($4-8\unit{GHz}$). Low-pass filters are used to minimize the parasitic noise coming down the dc bias line and attenuators are placed on the ac bias line to dampen noise generated by room-temperature electronics. The signal generated by the junction in the $4-8\unit{GHz}$ range goes through two circulators, used to isolate the sample from the amplification line noise, and is then amplified by a high electron mobility transistor amplifier placed at $3\unit{K}$.

At room temperature, the signal is separated in two branches, each entering an IQ mixer. One mixer is referenced to a local oscillator at frequency $f_1$ and the other at frequency $f_2=f_0-f_1$; for single-mode squeezing experiments, only one IQ mixer is used. All three microwave sources at frequencies $f_0$, $f_1$ and $f_2$ are phase coherent. The two IQ mixers take the signal emitted by the junction and provide the in-phase $X_{1,2}$ and quadrature $P_{1,2}$ parts relative to their references with a bandwidth of $80\unit{MHz}$. Similar set-ups have already been used to determine statistical properties of radiation in the microwave domain \cite{Menzel2010,Mariantoni2010,Menzel2012,Bozyigit2011,Eichler2011}. 

Any two quantities $A,B$ among $X_1$, $X_2$, $P_1$ and $P_2$ can be digitized simultaneously by a two-channel digitizer at a rate of $400\unit{MS/s}$, yielding a 2D probability map $\mathcal{P}\inp{A,B}$.  From $\mathcal{P}\inp{A,B}$, one can calculate any statistical quantity, in particular the variances $\moy{A^2}$, $\moy{B^2}$ $\moy{\inp{A^2+B^2}/2}$, as well as the correlators $\moy{AB}$. For single-mode squeezing experiments, we worked at $f_1=7\unit{GHz}$ and either $f_0=f_1$ or $f_0=2f_1$. For two-mode squeezing experiments, we chose to work at $f_0=14.5\unit{GHz}$, $f_1=7\unit{GHz}\Rightarrow f_2=7.5\unit{GHz}$.

\subsection{Calibration}
The four detection channels must be calibrated separately. This is achieved by measuring the variances $\moy{X_{1,2}^2}$, $\moy{P_{1,2}^2}$ with $\vac=0$. These should all be proportional to the noise spectral density of a tunnel junction at frequency $f_{1,2}$, given by $S\inp{\vdc,f}=\insb{S_0\inp{f+e\vdc/h}+S_0\inp{f-e\vdc/h}}/2$ where $S_0\inp{f}=\inp{hf/R}\coth\inp{hf/2k_BT}$ is the equilibrium noise spectral density at frequency $f$ in a tunnel junction of resistance $R$. When $\vdc=\vac=0$, only vacuum fluctuations are responsible for the observed noise spectral density: $S_{\mathrm{vacuum}}=S_0\inp{f}_{T=0}=hf/R$. By fitting the measurements with this formula, we find an electron temperature of $T=18\unit{mK}$ and an amplifier noise temperature of $\sim 3\unit{K}$, identical for all four channels. The small channel cross-talk is eliminated using the fact that $\moy{A_1B_2}=0$ when no microwave excitation is present.

In the presence of an ac excitation at frequency $f_0$ and amplitude $\vac$, the spectral density of current fluctuations $\widetilde{S}(\vdc,f,\vac,f_0)$ is given by:

\begin{equation}
\widetilde{S}\inp{\vdc,f,\vac,f_0}=\sum_{n=-\infty}^\infty J_n^2\inp{\frac{e\vac}{hf_0}}
S\inp{V_{dc}+n\frac{hf_0}{e},f}
\label{eq:Sp}
\end{equation}%
with $J_n$, the Bessel functions of the first kind. This quantity, the so-called photo-assisted noise, has been first predicted in \cite{Lesovik1994} and observed in \cite{Schoelkopf1998,Kozhevnikov2000}. From the measurement of $\widetilde{S}$, we can calibrate the excitation power, i.e. know what $\vac$ is experienced by the sample. 

\begin{figure}
    \centerline{ \includegraphics[width=\columnwidth] {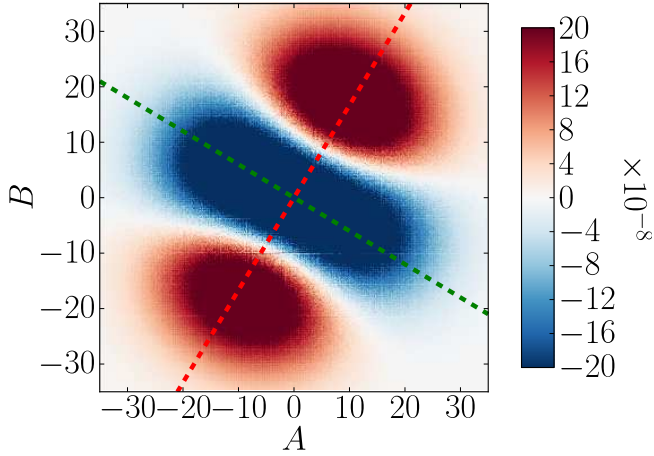}}
    \caption{\footnotesize (colour online) $\Delta \mathcal{P}= \mathcal{P}\inp{A,B}_{\vdc,\vac, f_0}- \mathcal{P}\inp{A,B}_{0,0}$ (unitless): Difference between the normalized 2D current fluctuation distributions of the tunnel junction at $f_1=7\unit{GHz}$, $\vdc=-26.6\unit{\mu V}\simeq hf_1/e$, $18\unit{mK}$, under $f_0=14\unit{GHz}$, $\vac=37\unit{\mu V}$ microwave excitation and without excitation at $\vdc=0$. $A$ and $B$ represent combinations of quadratures $X_1$ and $P_1$ of the observed signal (see text for details). Each distribution is made up of $2\ee{11}\unit{samples}$. Data was normalized so that the integral of $\mathcal P$ equals 1./(sans unit\'es) Diff\'erence entre les distributions bidimensionnelles normalis\'ees des fluctuations de courant de la jonction tunnel \`a $f_1=7\unit{GHz}$, $\vdc=-26,6\unit{\mu V}\simeq hf_1/e$, $18\unit{mK}$, sous excitation micro-ondes de $f_0=14\unit{GHz}$, $\vac=37\unit{\mu V}$  et sans excitation \`a $\vdc=0$. $A$ et $B$ repr\'esentent des combinaisons des quadratures $X_1$ et $P_1$ du signal observ\'e (voir le texte pour les d\'etails). Chaque distribution est constitu\'ee de $2\ee{11}\unit{\acute echantillons}$. Les donn\'ees ont \'et\'e normalis\'ees afin que l'int\'egrale de $\mathcal P$ \'egale 1.}
    \label{oneModeMap}
\end{figure}

\begin{figure}
    \centerline{ \includegraphics[width=\columnwidth] {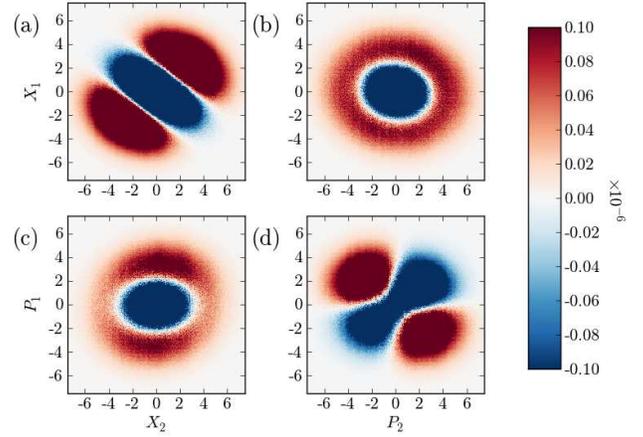}}
    \caption{\footnotesize (colour online) $\Delta \mathcal{P}= \mathcal{P}\inp{A,B}_{\vdc,\vac,f_0}- \mathcal{P}\inp{A,B}_{0,0}$ (unitless): (a)-(d) Difference between the normalized 2D current fluctuation distributions of the tunnel junction at $\vdc=29.4\unit{\mu V}$, $18\unit{mK}$, under $f_0=14.5\unit{GHz}$, $\vac=37\unit{\mu V}$ microwave excitation and without excitation at $\vdc=0$. $X$ and $P$ represent 2 quadratures of the observed signal while numbers 1 and 2 represent fluctuations of frequency $7\unit{GHz}$ and $7.5\unit{GHz}$ in arbitrary units. Each distribution is made up of $2\ee{11}\unit{samples}$. Data was normalized so that the integral of $\mathcal P$ equals 1./(sans unit\'es) (a)-(d) Diff\'erence entre les distributions bidimensionnelles normalis\'ees des fluctuations de courant de la jonction tunnel \`a $\vdc=29,4\unit{\mu V}$, $18\unit{mK}$, sous excitation micro-ondes de $f_0=14,5\unit{GHz}$, $\vac=37\unit{\mu V}$  et sans excitation \`a $\vdc=0$. $X$ et $P$ repr\'esentent deux quadratures du signal observ\'e tandis que les indices $1$ et $2$ repr\'esentent les fluctuations de fr\'equences $7\unit{GHz}$ et $7,5\unit{GHz}$ en unit\'es arbitraires. Chaque distribution est constitu\'ee de $2\ee{11}\unit{\acute echantillons}$. Les donn\'ees ont \'et\'e normalis\'ees afin que l'int\'egrale de $\mathcal P$ \'egale 1.\frPerm}
    \label{colMap}
\end{figure}

The IQ mixer outputs return $I_{1,2}=X_{1,2}\cos\theta+P_{1,2}\sin\theta$ and $Q_{1,2}=X_{1,2}\sin\theta-P_{1,2}\cos\theta$. Rather than controlling the phase $\theta$ of the incoming signal, the obtained results are plotted as 2D probability maps for each $\vdc$, both under ac excitation and at $\vac=0$. A map $\mathcal P\inp{A,B}_{0,0}$ taken at $\vdc=\vac=0$ represents the contribution of the vacuum fluctuations generated by the sample plus that of the amplifier, the latter dominating the signal. It follows that all maps differ only slightly from a 2D Gaussian distribution. To enhance the effect of the bias and excitation voltages, the difference $\mathcal P\inp{A,B}_{\vdc,\vac}-\mathcal P\inp{A,B}_{0,0}$ is plotted on Figs. \ref{oneModeMap} and \ref{colMap}, in which blue areas show regions where measured noise level is lower than that of the vacuum state while red zones indicate a higher-than-vacuum noise level. 

Fig. \ref{oneModeMap} represents single-mode squeezing at $f_0=2f_1$. It should be quite obvious that the symmetry axes of the distribution aren't aligned with the vertical and horizontal axes of the map, meaning the phase $\theta\neq n\pi/2$. The data can still be analysed in terms of $X_1$ and $P_1$ following a simple rotation of $-\theta$ on the data, which would align $X_1$ and $P_1$ along the red and green dashed lines added on this figure. Fig. \ref{colMap} shows 2-mode squeezing between $f_1 = 7\unit{GHz}$ and $f_2=7.5\unit{GHz}$. The phase appears here to be $\theta\simeq\pi/4$; however, this is misleading given that the signals represented in channels $A$ and $B$ have respective frequencies of $7\unit{GHz}$ and $7.5\unit{GHz}$, meaning their phases relative to the $f_0$ signal generator are not necessarily $\theta_1=\theta_2=\theta=0$. But since the only expected correlations should be found between $X_1$ and $X_2$ or between $P_1$ and $P_2$, any meaningful contribution of $P_1$ to channel $I_1$ ($X_1$ to channel $Q_1$) would show up in the $\mathcal P\inp{X_1,P_2}$ ($\mathcal P\inp{P_1,X_2}$) maps (subfigures (b) and (c)), where no squeezing is apparent. A similar argument can be made for contributions $X_2$ and $P_2$ to channels $Q_2$ and $I_2$. It follows that the relative phase between channels $A$ and $B$ is irrelevant so long as their respective phases relative to $f_0$ are consistent. 

\section[Single-mode squeezing]{Single-mode squeezing \cite{fn2}}
\subsection{Theory}
We measure the amplitudes $X$ and $P$ of the two quadratures of the electromagnetic field generated by the tunnel junction at a given frequency $f_1$ or $f_2$\footnote{subscripts have been removed to simplify notation}. According to recent work\cite{Grimsmo2016}, the corresponding quantum operators $\hat{X}$ and $\hat{P}$ are related to the electron current operator at frequency $f$, $\hat{I}(f)$ by:

\begin{equation}
\hat{X}=\frac{1}{\sqrt{2}} \inp{\hat{I}\inp{f})+\hat{I}^\dagger\inp{f}},\;\;
\hat{P}=\frac{i}{\sqrt{2}} \inp{\hat{I}\inp{f})-\hat{I}^\dagger\inp{f}}
\label{eqXP}
\end{equation}%
with $\hat{X}^\dagger=\hat{X}$, $\hat{P}^\dagger=\hat{P}$ and $\hat{I}^\dagger\inp f=\hat{I}\inp{-f}$. This allows us to relate the quadratures to properties of the electrons crossing the junction. The average commutator of those two observables  $\moy{\insb{\hat{X},\hat{P}}} = i \moy{\insb{\hat{I}\inp f,\hat{I}\inp {-f }}}=iS_{\mathrm{vacuum}}\inp f$ is non-zero, so uncertainties in the measurement of $X$ and $P$ obey Heisenberg's uncertainty principle: $\Delta X^2\Delta P^2\geq S_{\mathrm{vacuum}}^2$ with $\Delta O^2= \moy{\inp{\hat{O}-\moy{\hat{O}}}^2}$ for $\hat{O}=\hat{X},\hat{P}$. The following computations are made for $\moy{\hat{O}}=0$, meaning $\Delta O^2= \moy{\hat{O}^2}$. These variances are related to current-current correlators:

\begin{widetext}
\begin{equation}
\begin{array}{rl}
\Delta X^2&\!=\frac12\moy{\incb{\hat I\inp f,\hat I\inp{-f}}}+\frac12\insb{\moy{I\inp f^2}+\moy{I\inp{-f}^2}}=\widetilde{S}+\mathcal X \\[\bigskipamount]
\Delta P^2&\!=\frac12\moy{\incb{\hat I\inp f,\hat I\inp{-f}}}-\frac12\insb{\moy{I\inp f^2}+\moy{I\inp{-f}^2}}=\widetilde{S}-\mathcal X,
\label{eqDeltas}
\end{array}
\end{equation}
\end{widetext}
where the anti-commutator $\widetilde{S}=\frac{\moy{\incb{\hat I(f),\hat I(-f)}}}{2}$ is the usual noise and $\mathcal{X}=\frac{\insb{\moy{I\inp f^2}+\moy{ I\inp{-f}^2}}}{2}$ the correlator describing the noise dynamics, studied in \cite{Gabelli2008,Gabelli}, which is non-zero only if $2f=mf_0$ with $m$, an integer. When $\vac=0$, $\mathcal X=0$ and $\Delta X^2=\Delta P^2=\widetilde{S}$, which corresponds to $S_{\mathrm{vacuum}}$ at $\vdc=0$. The condition for single-mode squeezing is thus $\widetilde{S}\pm\mathcal{X} < S_{\mathrm{vacuum}}$. $\widetilde{S}$ is given by Eq. (\ref{eq:Sp}) while 

\begin{equation}
\mathcal X=\frac12\inp{\mathcal X_m\inp{f,f_0}+\mathcal X_{-m}\inp{-f,f_0}}
\label{Xsym}
\end{equation}

with 

\begin{equation}
\begin{array}{rl}
\mathcal X_m\inp{f,f_0}&=\moy{\hat{I}\inp{f}\hat{I}\inp{mf_0-f}}\\
                       &=\sum_n \frac{\alpha_n}{2}[S_0\inp{f_{n+}}+\inp{-1}^mS_0\inp{f_{n-}}],
\end{array}
\label{eqX}
\end{equation}
using $\alpha_n=J_n\inp{e\vac/hf_0} J_{n+m}\inp{e\vac/hf_0}$ and $f_{n\pm}=\vdc \pm h\inp{f+nf_0}/e$. The sum here is due to the interference which occurs when $n$ photons are absorbed and $n+m$ are emitted or vice-versa. Each of these absorptions and emissions are weighted by the $\alpha_n$ amplitudes. This is the basis for all correlator theoretical predictions presented here\cite{Gabelli2008,Gabelli,Gabelli2007}. To simplify the discussion, we introduce here the dimensionless operators

\begin{equation}
\hat{x}=\frac{\hat{X}}{\sqrt{2hf/R}},\;\;\hat{p}=\frac{\hat{P}}{\sqrt{2hf/R}},
\label{eqPxp}
\end{equation}
chosen so that $\moy{\insb{\hat x, \hat p }}=i$. Thus, vacuum fluctuations correspond to $\moy{x^2}=\moy{p^2}=1/2$, the Heisenberg uncertainty relation to $\moy{x^2}\moy{p^2}\geq1/4$ and squeezing to $\moy{x^2}<1/2$ or $\moy{p^2}<1/2$. It follows that $\moy{x^2+p^2}/2=\widetilde S/S_{\mathrm{vacuum}}$.

\subsection{Experimental Results}
\begin{figure}
    \centerline{ \includegraphics[width=\columnwidth] {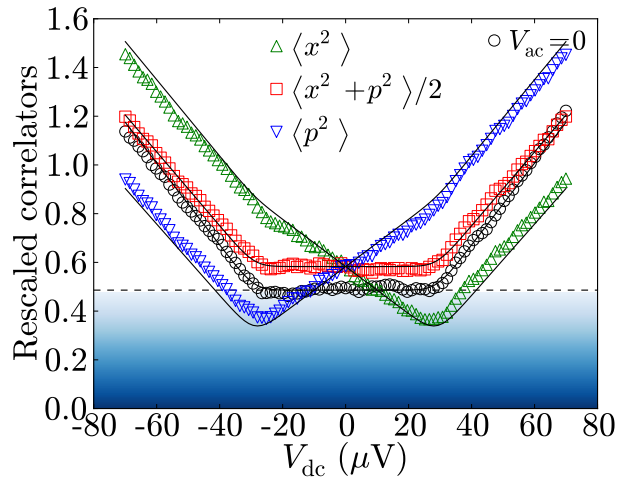}}
    \caption{\footnotesize (colour online)  Rescaled (unitless) variances of the EM field generated by a $70\unit{\Omega}$ tunnel junction at $18\unit{mK}$ under $f_0=14\unit{GHz}$, $\vac=37\unit{\mu V}$ microwave excitation and without excitation, obtained using signal quadratures at frequency $f_1=7\unit{GHz}$. Symbols represent experimental data, with symbol sizes representing experimental uncertainty, and lines are theoretical expectations of Eqs. \ref{eq:Sp} and \ref{Xsym}. The shaded area showcases the less-than-vacuum noise levels./Variances remises \`a l'\'echelle (sans unit\'es) du champ \'electromagn\'etique g\'en\'er\'e par une jonction tunnel de $70\unit{\Omega}$ sous excitation micro-ondes de $f_0=14\unit{GHz}$, $\vac=37\unit{\mu V}$ et sans excitation, obtenues \`a partir des quadratures de signal \`a la  fr\'equence $f_1=7\unit{GHz}$. Les symboles repr\'esentent les r\'esultats exp\'erimentaux et les lignes, les attentes th\'eoriques des \'Eqs. \ref{eq:Sp} et \ref{Xsym}. La zone en d\'egrad\'e indique les niveaux de bruit inf\'erieurs au niveau du vide.}
    \label{harm2}
\end{figure}

\begin{figure}
    \centerline{\includegraphics[width=\columnwidth] {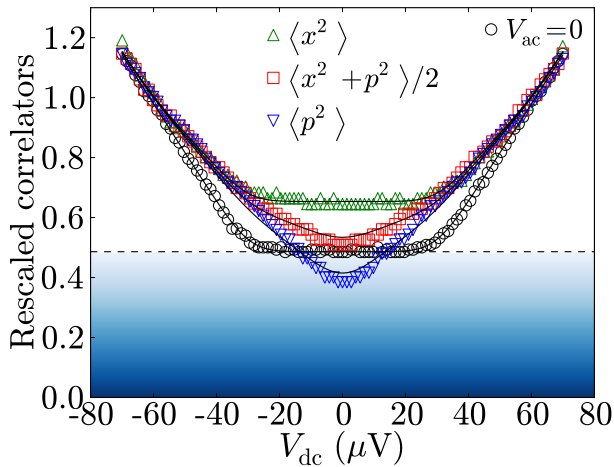}}
    \caption{\footnotesize (colour online)  Rescaled (unitless) variances of the EM field generated by a $70\unit{\Omega}$ tunnel junction at $28\unit{mK}$ under $f_0=7.2\unit{GHz}$, $\vac=36\unit{\mu V}$ microwave excitation and without excitation, obtained using signal quadratures at frequency $f_1\simeq7.2\unit{GHz}$. Symbols represent experimental data and lines are theoretical expectations of Eqs. \ref{eq:Sp} and \ref{Xsym}. The shaded area showcases the less-than-vacuum noise levels. Data adapted from Ref. \cite{Gasse2013}, Fig. 3./Variances remises \`a l'\'echelle (sans unit\'es) du champ \'electromagn\'etique g\'en\'er\'e par une jonction tunnel de $70\unit{\Omega}$ \`a $28\unit{mK}$ sous excitation micro-ondes de $f_0=7,2\unit{GHz}$, $\vac=36\unit{\mu V}$ et sans excitation, obtenues \`a partir des quadratures de signal \`a la  fr\'equence $f_1\simeq7,2\unit{GHz}$. Les symboles repr\'esentent les r\'esultats exp\'erimentaux et les lignes, les attentes th\'eoriques des \'Eqs. \ref{eq:Sp} et \ref{Xsym}. La zone en d\'egrad\'e indique les niveaux de bruit inf\'erieurs au niveau du vide. Donn\'ees adapt\'ees de la Fig. 3 de la R\'ef. \cite{Gasse2013}.}
    \label{harm1}
\end{figure}

We first consider the noise measured when the junction in not under ac excitation. In this case, nothing sets an absolute phase in the measurement, so $\Delta X^2=\Delta P^2$, as shown on Figs. \ref{harm2} and \ref{harm1} (black circles). This experiment is equivalent to the usual variance measurement of the sample current fluctuations using a power detector, i.e. $\moy{X^2}=\Delta X^2=\moy{P^2}=\Delta P^2 = G\inp{S_{amp}+S\inp{\vdc,f_1}}$ with $G$ the gain of the setup, $S_{amp}$ the current noise spectral density of the amplifier and $S\inp{\vdc,f_1}$ the noise spectral density. $S\inp{\vdc,f_1}$ is constant as long as $eV<hf_1$, as showcased by the wide plateau around $\vdc=0$ on Figs. \ref{harm2} and \ref{harm1}.

We now turn to the case where excitation and detection are synchronized, i.e. we perform phase sensitive noise measurements. Eqs.~\ref{eqDeltas}--\ref{eqX} can be used as theoretical predictions for all correlators. The corresponding values are rescaled and represented as black lines along with rescaled experimental results on Fig. \ref{harm2} for an excitation at frequency $f_0=2f_1$ and on Fig. \ref{harm1} for $f_0=f_1$. As expected, $\moy{x^2+p^2}/2$ correspond to the photo-assisted shot noise introduced in Eq. \ref{eq:Sp}. This matches perfectly with experimental data (red squares on Figs. \ref{harm2} and \ref{harm1}).

The optimal conditions for the observation of squeezing are very different for $m=1$ ($f_0=2f_1$, which corresponds to four-wave mixing) and $m=2$ ($f_0=f_1$, which corresponds to three-wave mixing). These can be easily understood at $T=0$, when $\widetilde{S}$ is a piecewise linear function. For $m=1$, $\widetilde{S}$ is independent of $\vdc$ as long as $\vdc<hf$, while $\mathcal X$ is a linear function of $\vdc$. As a result, the optimal $\vdc$ is $hf_1/e$ (see Fig. \ref{harm2}). We find that the maximal squeezing at $T=0$, $m=1$ corresponds to $\Delta X^2=0.62S_{\mathrm{vacuum}}$, i.e. $2.09\unit{dB}$ below vacuum. This corresponds to four-wave mixing. For $m=2$, $\widetilde{S}$ is minimal at $\vdc=0$ and increases as $\abs{\vdc}$ while $\mathcal X$ is maximal at $\vdc=0$ and decreases following $-\abs{\vdc}$. Thus the optimal squeezing occurs at $\vdc=0$. We find that the maximal squeezing at $T=0$, $m=2$ corresponds to $\Delta X^2=0.73S_{\mathrm{vacuum}}$, i.e. $1.37\unit{dB}$ below vacuum, see Fig. \ref{harm1}. These results are independent of $f_1$ at zero temperature. 

In both cases, agreement between theory and experiment is very good.
There is a range of $\vdc$ where one quadrature is below the plateau corresponding to vacuum fluctuations: this corresponds to squeezing of the electromagnetic field generated by the junction. We have performed such measurements for many excitation powers. The optimal squeezing for $f_0=2f_1$ (Fig. \ref{harm2}) is found to occur at $\vdc\simeq hf_1/e$ and corresponds to $0.74$ times vacuum fluctuations, i.e. $1.31\unit{dB}$ below vacuum. The optimal squeezing for $f_0=f_1$ (Fig. \ref{harm1}) occurs at $\vdc=0$ and corresponds to $0.82$ times vacuum fluctuations, i.e. $0.86\unit{dB}$ below vacuum. 

The existence of such squeezing has been predicted recently \cite{Bednorz2013} and is a particular case of a Cauchy-Schwarz inequality violation by electronic quantum noise. It is remarkable to note that the squeezing comes from the $\mathcal X$ term, which reflects the modulation of the shot noise by a time-dependent voltage. The origin of this term is in the shot noise itself, which comes from the granularity of the charge.

\section{Two-mode squeezing}
\subsection{Theory}
While single-mode squeezing can be simply defined using the two dimensionless operators of Eq. \ref{eqPxp}, two-mode squeezing requires a slightly more complex description. We work here exclusively at $m=1$, $f_0=f_1+f_2$. Since the quadratures are studied at different frequencies, no single-mode squeezing should be observed. Therefore, $\moy{\hat I^2\inp{f_k}}=0$ and $\moy{X_k^2}=\moy{P_k^2}=\widetilde{S}\inp{f_k}$. Therefore, new operators are needed that combine those of Eq. \ref{eqPxp} for squeezing to be observed. These are 

\begin{equation}
\begin{array}{rlr}
\hat{u}\!&=\frac{\hat{x}_1-\hat{x}_2}{\sqrt{2}}; &\;\;\;\;\hat{u}'=\frac{\hat{x}_1+\hat{x}_2}{\sqrt{2}};\\[\bigskipamount]
\hat{v}\!&=\frac{\hat{p}_1+\hat{p}_2}{\sqrt{2}};
&\;\;\;\;\hat{v'}=\frac{\hat{p}_1-\hat{p}_2}{\sqrt{2}}.
\label{equv}
\end{array}
\end{equation}
These new operators redefine squeezing as the possibility of either $\moy{\hat{u}^2}$, $\moy{\hat{v}^2}$,$\moy{\hat{u}'^2}$ or $\moy{\hat{v}'^2}$ going below vacuum fluctuation levels. As will be shown on Fig. \ref{I1I2Q1Q2}, $\moy{x_kp_{k'}}=0\;\;\forall\;\;k$. It follows that operators of the type $\inp{x_1±p_2}/\sqrt2$ and $\inp{p_1±x_2}/\sqrt2$ would reduce to variances $\inp{\moy{x_1^2}+\moy{p_1^2}}/2=\inp{\moy{x_1^1}+\moy{p_2^2}}/2$, which are always above vacuum. These redefinitions are therefore not necessary here.

\begin{figure}
    \centerline{\includegraphics[width=\columnwidth]{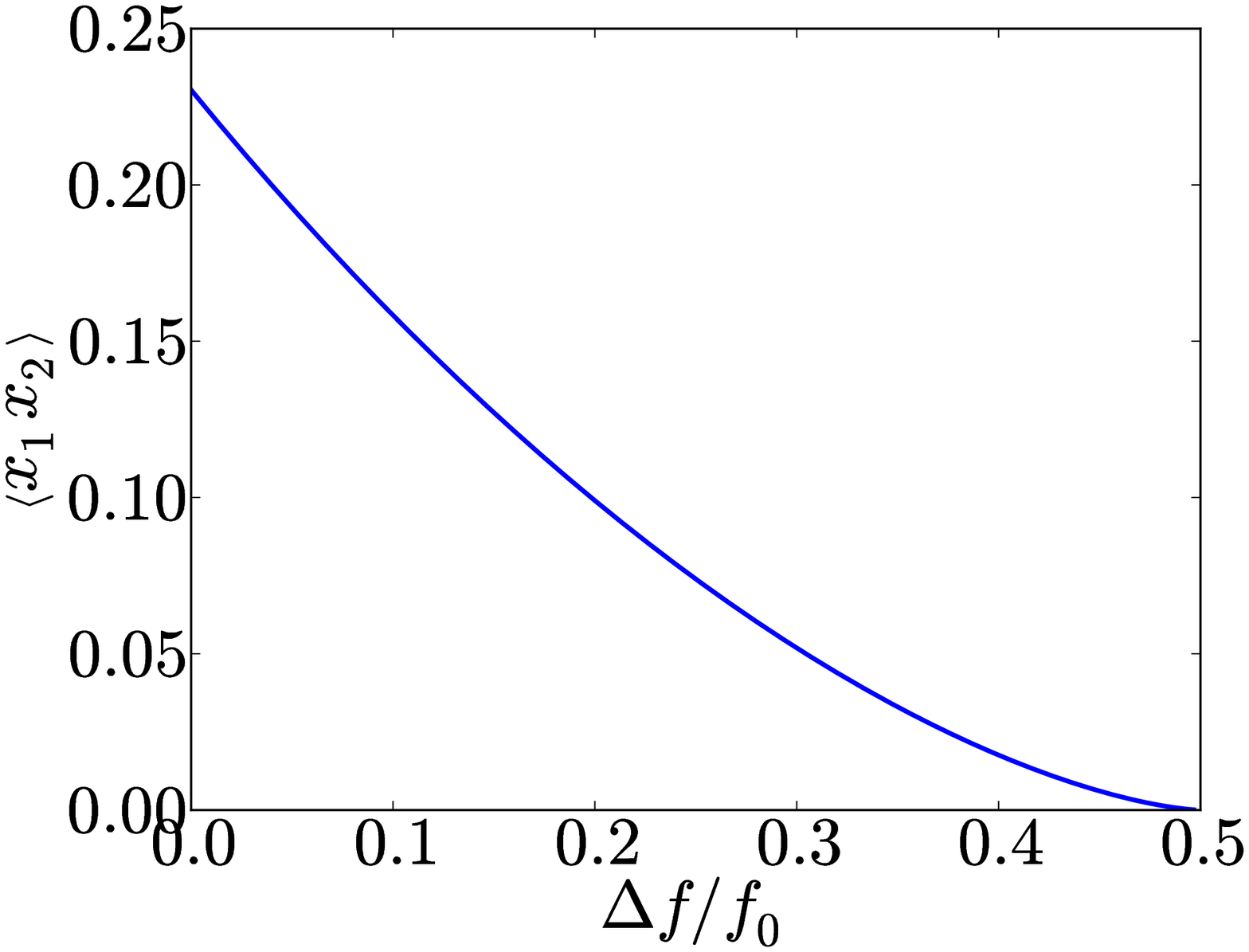}}
		\centerline{\includegraphics[width=\columnwidth] {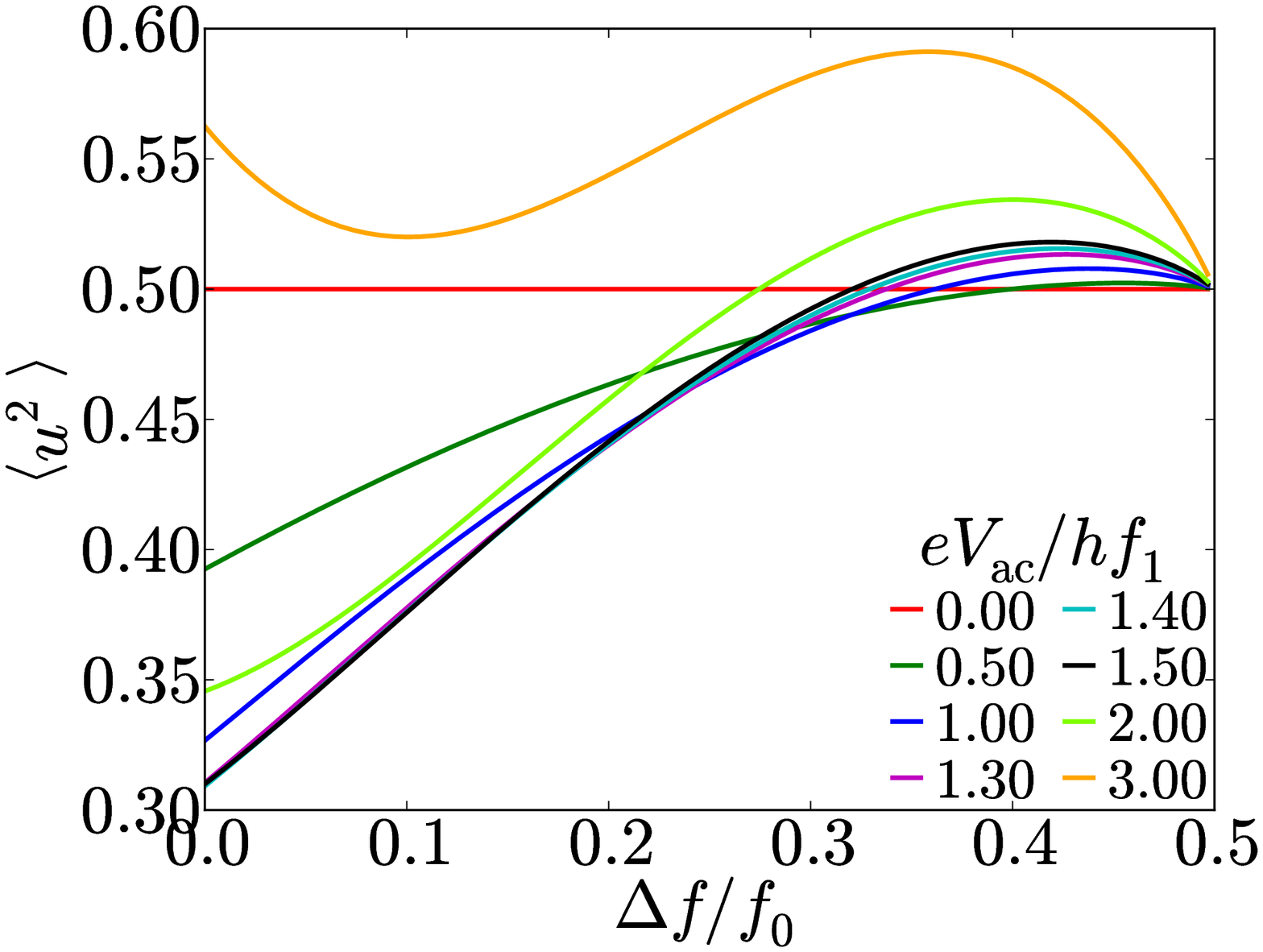}}
    \caption{\footnotesize (top) $\moy{x_1x_2}$ as a function of $\Delta f/f_0$ with $f_1=f_0/2-\Delta f$, $f_2=f_0/2+\Delta f$ at $\vdc=hf_1/e$, $\vac=1.5hf_1/e$, $T=0\unit K$ where $x_{1,2}$ is given by Eq. \ref{eqPxp}. (bottom) Search for optimal $\vac$ for two-mode squeezing observation (lowest point under 0.5) at $\vdc=hf_1/e$./ (haut) $\moy{x_1x_2}$ en fonction de $\Delta f/f_0$ avec $f_1=f_0/2-\Delta f$, $f_2=f_0/2+\Delta f$ \`a $\vdc=hf_1/e$, $\vac=1.5hf_1/e$, $T=0\unit K$, o\`u $x_{1,2}$ est donn\'e par Eq. \ref{eqPxp}. (bas) D\'etermination de la tension ac optimale pour l'observation de la compression d'\'etats \`a deux modes maximale (point le plus bas sous 0,5) \`a $\vdc=hf_1/e$.}
    \label{figFopt}
\end{figure}

If $\vac=0$, there can be no correlation between the currents observed at two different frequencies, meaning $ \moy{\hat{I}\inp{\pm f_1} \hat{I}\inp{\pm f_2}}=0$. However, an excitation of finite amplitude at frequency $f_0=f_1+f_2$ induces correlations $\moy{\hat{I}\inp{f_1}\hat{I}\inp{f_2}}= \moy{\hat{I}\inp{-f_1}\hat{I}\inp{-f_2}}=\mathcal X_{m=1}\inp{f_1,f_0}$. It follows that $\moy{X_1X_2}=-\moy{P_1P_2}=\mathcal X$. The proper choice of frequencies is crucial to optimizing this value. First, quantum effects are prominent at high frequencies, i.e. $f_{1,2}\gg k_BT/h$. Here, these frequencies are limited by our $4-8\unit{GHz}$ cryogenic electronics. Moreover, as can be seen on the left side of Fig. \ref{figFopt}, a lower $\Delta f=f_0/2-f_1=f_2-f_0/2$ leads to a greater value for $\mathcal X=\moy{X_1X_2}$. However, $f_1$ and $f_2$ must be sufficiently far apart for mixers to distinguish between them; for the experiments presented here, $\Delta f = 0.25\unit{GHz}$. The right panel of Fig. \ref{figFopt} can be used to ascertain the ideal $\vac$ for observing two-mode squeezing. Based on that analysis, the optimal values can be observed at $\vdc\simeq hf_1/e$, $\vac\simeq 1.5hf_1/e$ using $f_1=7\unit{GHz}$, $f_2=7.5\unit{GHz}$ and $f_0=14.5\unit{GHz}$.

\subsection{Experimental Results}
\begin{figure}
    \centerline{ \includegraphics[width=\columnwidth] {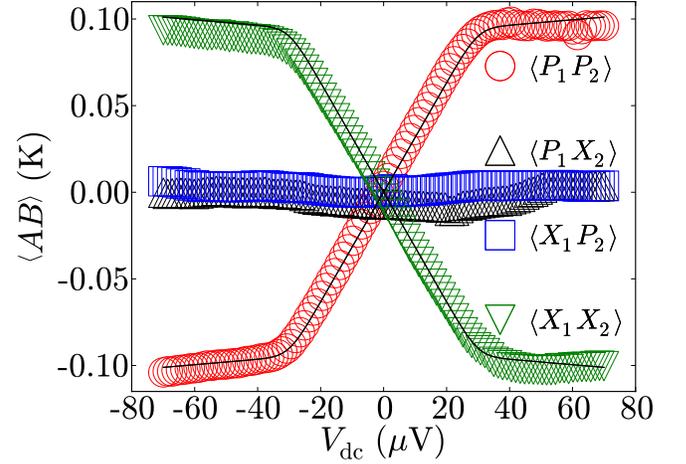}}
    \caption{\footnotesize (colour online) Quadrature correlators of EM field at frequencies $f_1=7\unit{GHz}$, $f_2=7.5\unit{GHz}$ generated by a $70\unit{\Omega}$ tunnel junction at $18\unit{mK}$ under $f_0=14.5\unit{GHz}$, $\vac=37\unit{\mu V}$ microwave excitation. Symbols represent experimental data, with symbol sizes representing experimental uncertainty. Lines are theoretical expectations based on Eq. \ref{Xsym}. /Corr\'elateurs de quadrature du champ \'electromagn\'etique aux fr\'equences $f_1=7\unit{GHz}$, $f_2=7.5\unit{GHz}$ g\'en\'er\'e par une jonction tunnel de $70\unit\Omega$ \`a $18\unit{mK}$ sous excitation micro-ondes de $f_0=14,5\unit{Ghz}$, $\vac=37\unit{\mu V}$. Les symboles repr\'esentent les r\'esultats exp\'erimentaux et les lignes, les attentes th\'eoriques de l'\'Eq. \ref{Xsym}. \frPerm}
    \label{I1I2Q1Q2}
\end{figure}

Figs. \ref{colMap}(b) and (c), which correspond to $\Delta \mathcal{P}\inp{X_1,P_2}$ and $\Delta \mathcal{P}\inp{P_1,X_2}$, are almost invariant by rotation. This means that the corresponding probability $\mathcal{P}\inp{X,P}$ depends only on $X^2+P^2$. As an immediate consequence, one expects $\moy{X_1P_2}=\moy{P_1X_2}=0$: $X_1$ and $P_2$ are uncorrelated, as are $X_2$ and $P_1$. In contrast, Figs. \ref{colMap}(a) and (d), which show respectively $\Delta \mathcal{P}\inp{X_1,X_2}$ and $\Delta \mathcal{P}\inp{P_1,P_2}$, are not invariant by rotation: the axes $X_1=\pm X_2$ ($P_1=\pm P_2$) are singular: for a given value of $X_1$($P_1$) the probability $\Delta \mathcal{P}\inp{X_1,X_2}$ ($\Delta \mathcal{P}\inp{P_1,P_2}$) is either maximal or minimal for $X_2=X_1$ ($P_2=-P_1$). This demonstrates the possibility of observing correlations or anticorrelations between $X_1$ and $X_2$ on one hand and between $P_1$ and $P_2$ on the other. Data in Figs. \ref{colMap}(a) through (d) correspond to two frequencies $f_1$ and $f_2$ that sum up to $f_0$, all three frequencies being phase coherent. If this condition is not fulfilled, no correlations are observed between any two quadratures, giving plots similar to Figs. \ref{colMap}(b) or (c) (data not shown). The effect of frequencies on correlations between power fluctuations has been thoroughly studied in Refs. \cite{C4classique,Forgues2014}. To be more quantitative, we show on Fig. \ref{I1I2Q1Q2} the $\moy{AB}$ correlators as a function of the dc bias voltage for a fixed $\vac$. Clearly, $\moy{X_1P_2}=\moy{P_1X_2}=0$ while $\moy{X_1X_2}=-\moy{P_1P_2}$ is non-zero for $\vdc\neq0$. These results are presented in temperature units (K), using the usual unit conversion $T_{noise}=RS/2k_B$ for the measured noise spectral density $S$ of a conductor of resistance $R$. Using Eq. \ref{Xsym}, this can be represented theoretically by $-\moy{P_1P_2}=\moy{X_1X_2}=\mathcal X$, which once again fits the experimental data very well.

Once again, at $k_BT\ll hf_{1,2}$ and $\vdc=\vac=0$, the noise emitted by the junction is equivalent to vacuum fluctuations, which can be seen experimentally as a plateau at low $\vdc$ on Fig. \ref{mixData} in $\moy{\hat{x}_k^2}$ vs $\vdc$ at $\vac=0$ (black circles). This vacuum noise level is outlined by a dashed line, the shaded area indicating less-than-vacuum noise levels. 

Fig. \ref{mixData} illustrates clearly the lack of single-mode squeezing for $f_0=f_1+f_2$ when $f_1\neq f_2$ since $\moy{\hat{x}_1^2}>1/2$\footnote{as well as $\moy{\hat{x}_2^2}=\moy{\hat{p}_1^2}=\moy{\hat{p}_2^2}>1/2$, data not shown}. It it obvious on Fig. \ref{mixData} that $\moy{\hat{u}^2}\simeq\moy{\hat{v}^2}$ goes below $1/2$ for certain values of  $\vdc$. This proves that two-mode squeezing can be observed in electronic shot noise. Once again, theoretical expectations can be plotted using Eq. \ref{Xsym}. These are shown as lines along experimental results after normalization on Fig. \ref{mixData}.

The optimal observed squeezing corresponds to $\moy{\hat{u}^2}=0.32\pm0.05$, $\moy{\hat{v}^2}=0.31\pm0.05$, i.e. $2.1\unit{dB}$ below vacuum, versus the theoretical expectation of $\moy{\hat{u}^2}=\moy{\hat{v}^2}=0.33$. This minimum is observed at  $\vdc\simeq30\unit{\mu V}\simeq hf_{1,2}/e$. All data are in good agreement with theoretical predictions, plotted as full black lines on Fig. \ref{mixData} with $\moy{\hat{x}_k^2}=\moy{\hat{p}_k^2}=\widetilde{S}\inp{f_k}$, the photo-assisted shot noise given by Eq. \ref{eqX}, using $m=1$\cite{Lesovik1994}. Curves for $\moy{u'^2}$ and $\moy{v'^2}$ follow the same behaviour with reversed dc bias, showing minima of $0.35\pm0.05$ and $0.41\pm0.05$ or $1.6\unit{dB}$ at $\vdc\simeq -30\unit{\mu V}\simeq -hf_{1,2}/e$. The latter data was omitted from Fig. \ref{mixData} for simplicity.

\subsection{Entanglement}
\begin{figure}
    \centerline{\includegraphics[width=\columnwidth] {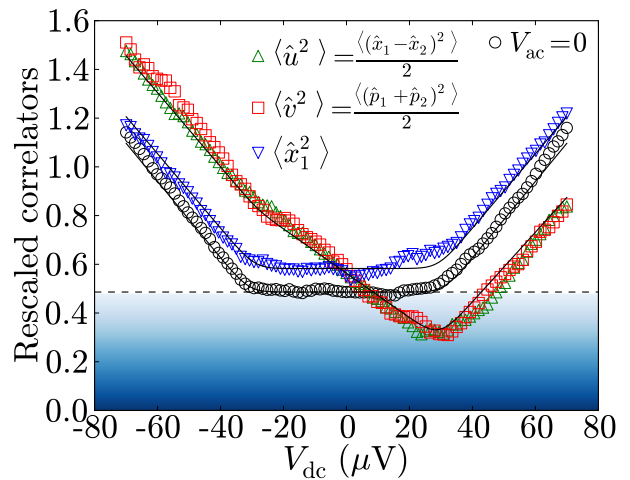}}
    \caption{\footnotesize (colour online)  Rescaled (unitless) variances of the EM field generated by a $70\unit{\Omega}$ tunnel junction at $18\unit{mK}$ under $14.5\unit{GHz}$, $\vac=37\unit{\mu V}$ microwave excitation and without excitation, obtained using signal quadratures at frequencies $f_1=7\unit{GHz}$, $f_2=7.5\unit{GHz}$. Symbols represent experimental data, with symbol sizes representing experimental uncertainty, and lines are theoretical expectations based on Eq. \ref{Xsym}. The shaded area showcases the less-than-vacuum noise levels./Variances remises \`a l'\'echelle (sans unit\'es) du champ \'electromagn\'etique g\'en\'er\'e par une jonction tunnel de $70\unit{\Omega}$ \`a $18\unit{mK}$ sous excitation micro-ondes de $f_0=14,5\unit{GHz}$, $\vac=37\unit{\mu V}$ et sans excitation, obtenues \`a partir des quadratures de signal aux fr\'equences $f_1=7\unit{GHz}$,  $f_2=7,5\unit{GHz}$. Les symboles repr\'esentent les r\'esultats exp\'erimentaux et les lignes, les attentes th\'eoriques de l'\'Eq. \ref{Xsym}. La zone en d\'egrad\'e indique les niveaux de bruit inf\'erieurs au niveau du vide. \frPerm}
    \label{mixData}
\end{figure}

While the presence of two-mode squeezing shows the existence of strong correlations between quadratures of the electromagnetic field at different frequencies, this is not enough to prove the existence of entanglement. A criterion certifying the inseparability of the two-modes, and thus entanglement between them, is given in terms of the quantity $\delta=\moy{\hat{u}^2}+\moy{\hat{v}^2}$. In the case of a classical field, this must obey $\delta>1$\cite{Duan2000}. This is equivalent to a Bell-like inequality for continuous variables. As we reported in Fig. \ref{mixData}, we observe $\delta=0.6\pm0.1$. Thus, photons emitted at frequencies $f_1$ and $f_2$ are not only correlated but also form EPR pairs suitable for quantum information processing with continuous variables\cite{Braunstein2005}.

Two-mode quadrature-squeezed states are usually characterized by their covariance matrix. Following the notations of Ref. \cite{Giedke2003}, our experiment corresponds to $n=2\moy{x_{1,2}^2}\simeq2\moy{p_{1,2}^2}$, $k=2\moy{x_1x_2}\simeq-2\moy{p_1p_2}$ so that $\delta=n-k$. Equilibrium at $T=0$ corresponds to $n=1$ and $k=0$. Our observed optimal squeezing corresponds to $n=1.3\pm0.1$ and $k=0.52\pm0.05$. From these numbers, one can calculate all the statistical properties that characterize the electromagnetic field generated by the junction.  In particular, we find a purity of $\mu=0.82$ (as defined in Ref. \cite{DiGuglielmo2007}), which would be 0 for a non-entangled state and 1 for a pure (maximally entangled) state. While in our experiment, the entangled photons are not spatially separated, this could easily be achieved using a diplexer, which can separate frequency bands without dissipation.

\section{Conclusion}

We have provided the first experimental demonstration of the existence of non-classical properties in the electromagnetic field radiated by a normal, non-superconducting conductor. More particularly, we have shown that the microwave radiation generated by a tunnel junction can exhibit squeezing and entanglement. Since the mechanism involved here is the ability to modulate the electron shot noise by an ac voltage, it is clear that these possibilities are offered by any coherent conductor and are not specific to the tunnel junction. Since the shot noise involves the Fano factor, it is probable that a diffusive wire or a QPC should generate less squeezing and entanglement, while a Normal-Superconductor structure might be an interesting system to explore.

We have excited the junction with a sine wave. It is probable that the use of a more clever periodic function with several harmonics \cite{Gabelli2013} might improve the degree of squeezing and entanglement of the emitted radiation. Moreover, the correlations induced by the ac excitation are extremely broadband, as can be seen on Fig. \ref{figFopt} (left). Thus, instead of considering, as we did here, the electromagnetic field, which is well defined in the frequency domain, it might be interesting to consider modes that embrace the full spectrum of squeezed radiation with adequate weighting, or in other words to consider the squeezing/entanglement in time domain.

In this paper, we have presented evidence of the quantum aspect of the electromagnetic field generated by a tunnel junction by measuring its quadratures. Other properties, like its photon statistics, should convey measurable quantum features. Indeed, the existence of correlations between photons of different frequencies at the single photon level have been reported \cite{Forgues2014}, associated with the existence of two-mode squeezing. Single-mode squeezing is also known to be related to the emission of photon pairs, which should show up in the photon statistics. Thanks to a recently developed link between the statistics of photons and that of quadratures of the field \cite{Virally2016}, and despite the absence of available photon counters in the microwave domains, the shot noise of photon pairs generated by electron shot noise has been observed \cite{Simoneau2016}. Further developments in the generation and detection of non-classical electromagnetic fields generated by quantum, but non superconducting, conductors will for sure appear in the forthcoming years.


\end{document}